\documentclass[aps,preprint,amsmath,amssymb,showpacs]{revtex4}

\usepackage{graphicx}
\begin{document}

\title{Anomalous WWH couplings in $\gamma\gamma$ collision with
initial beams and final state polarizations}

\author{Banu \c{S}ahin}
\email[]{bsahin@wisc.edu}\email[]{dilec@science.ankara.edu.tr}
\affiliation{Department of Physics, University of Wisconsin,
Madison, WI 53706, USA}
\affiliation{Department of Physics, Faculty
of Sciences, Ankara University, 06100 Tandogan, Ankara, Turkey}

\begin{abstract}
The constraints on anomalous Higgs boson couplings are investigated
through the process $\gamma\gamma \to W^{+}W^{-}H$. Considering the
longitudinal and transverse polarization states of the final $W^{+}$
and $W^{-}$ bosons and incoming beam polarizations, we find 95\%
confidence level limits on the anomalous coupling parameters $\Delta
a_{W}$, $b_{W}$ and $\beta_{W}$  with an integrated luminosity of
500 $fb^{-1}$ and $\sqrt{s}$=0.5, 1 TeV energies. We show that
initial beam and final state polarizations lead to a significant
improvement on the sensitivity limits of the anomalous coupling
parameters $b_{W}$ and $\beta_{W}$.

\end{abstract}


\maketitle

\section{Introduction}

The standard model (SM) of electroweak interactions based on the
gauge group $SU(2)_{L}\times U(1)_{Y}$ has been verified to be
successful in describing all the available precision experimental
data. The recent measurements of gauge boson couplings  at the CERN
$e^+e^-$ collider LEP and Fermilab Tevatron shed some light on the
correctness of SM predictions for gauge boson interactions. On the
other hand, the Higgs boson, which is a remnant of the spontaneous
breaking of gauge symmetry, is the only undiscovered ingredient of
the SM so far. The investigation of the electroweak symmetry
breaking mechanism and the search for the Higgs boson constitute the
prime targets of future colliders. Once the Higgs boson is found,
its properties and interactions with other particles may be studied
in detail with an $e^{+}e^{-}$ collider and its $\gamma\gamma$,
$e\gamma$ modes. However, the Higgs boson has not been discovered
yet, there are experimental bounds for its mass. A lower bound on
the Higgs mass is provided by direct searches at the LEP collider,
$m_{H}>114.4$ GeV \cite{barate}. Moreover, electroweak precision
measurements provide an upper bound on its mass, $m_{H}<186$ GeV
\cite{lep}.

In the literature there has been a great amount of work on Higgs
interactions with gauge bosons. Higgs production modes proceed via
its coupling with a pair of gauge bosons at a linear collider and
deviations from their SM values are probed via such production
processes. Anomalous WWH couplings have been investigated for the
process $e^{+}e^{-}\to f\bar{f}H$
\cite{biswal,garcia,barger,hagiwara}, $e^{+}e^{-}\to
W^{+}W^{-}\gamma$ \cite{lietti}, $e^{-}\gamma\to \nu_{e}W^{-}H$
\cite{choudhury} and $\gamma\gamma\to WWWW$ \cite{han}. Anomalous
gauge couplings of Higgs bosons have been analyzed at the LHC
through the weak boson scattering \cite{he} and vector boson fusion
\cite{hankele} processes.

At an $e^{+}e^{-}$ collision Higgs boson production processes most
often include both WWH and ZZH couplings and it is difficult to
dissociate WWH from ZZH \cite{biswal}. In this work we analyzed
anomalous WWH vertex via the process $\gamma\gamma \to W^{+}W^{-}H$.
This process isolates the WWH vertex and gives us the opportunity to
study the WWH vertex independent of the ZZH. Furthermore, with Higgs
and W bosons being visible (in their decay modes), one is offered a
large amount of kinematical variables in the construction of
suitable observables. Clearly, $e^{-}\gamma\to \nu_{e}W^{-}H$
process also isolates the WWH vertex as discussed in
Refs.\cite{choudhury}. It was shown that initial and final state
polarizations lead to a significant improvement in the sensitivity
limits of the anomalous coupling parameters $b_W$ and $\beta_W$.
However, we will show that the $\gamma\gamma \to W^{+}W^{-}H$
process at the $\gamma\gamma$ mode of a linear collider has a higher
potential to probe anomalous WWH couplings than $e^{-}\gamma\to
\nu_{e}W^{-}H$. We take into account initial beams and final W boson
polarizations to improve the sensitivity limits.

Deviations from SM expectations in the Higgs sector can be
parameterized in a model independent way by an effective Lagrangian.
We employ the effective lagrangian approach described in
Ref.\cite{biswal,garcia,barger,choudhury}. If we demand Lorentz
invariance and gauge invariance, the most general coupling structure
(retaining up to dimension six terms in the effective lagrangian)
can be expressed as

\begin{eqnarray}
\Gamma_{\mu\nu}^{V}=i\tilde{g_{V}}\left[a_{V}g_{\mu\nu}+\frac{b_{V}}{m_{V}^{2}}(k_{2\mu}k_{1\nu}-g_{\mu\nu}k_{1}.k_{2})
+\frac{\beta_{V}}{m_{V}^{2}}\epsilon_{\mu\nu\alpha\beta}k_{1}^{\alpha}k_{2}^{\beta}\right]
\end{eqnarray}
with
\begin{eqnarray}
&&\tilde{g}_{W}=g_{W}m_{W},\,\,\,\,\,\
\tilde{g}_{Z}=\frac{g_{W}m_{W}}{cos^{2}\theta_{W}}
\nonumber\\
&&g_{W}=\frac{g_{e}}{sin\theta_{W}},\,\,\,\,\,\
g_{e}=\sqrt{4\pi\alpha}
\end{eqnarray}
where $k_{1}^{\mu}$ and $k_{2}^{\mu}$ are the momenta of two W's (or
Z's). We consider that all the momenta are outgoing from the vertex.
In the context of the SM, at the tree level, couplings are given by
$a_{V}=1$, $b_{V}=0$ and $\beta_{V}=0$. In our calculations we
reparametrize the coupling $a_{V}$ as $a_{V}=\Delta a_{V}+1$,
therefore within the SM $\Delta a_{V}=0$.

\section{Polarized cross sections}

Experiments at future linear $e^{+}e^{-}$ colliders will be able to
investigate in detail the interactions of gauge bosons, fermions and
scalars. In particular, one of the prime targets is the study of the
interactions of the Higgs boson, for which the $\gamma\gamma$ mode
of the collider seems especially suitable
\cite{eboli,baillargeon,boos}.

The process $\gamma\gamma \to W^{+}W^{-}H$ takes part as a
subprocess in the $e^{+}e^{-}$ collision. Real gamma beams which
enter the subprocess are obtained through Compton backscattering of
laser light off an electron and a positron beam, where most of the
photons are produced at the high energy region. The luminosities for
$e\gamma$ and $\gamma\gamma$ collisions turn out to be of the same
order as that for $e^{+}e^{-}$ \cite{Ginzburg}, so the cross
sections for photoproduction processes with real photons are
considerably larger than the virtual photon case. The spectrum of
the backscattered photons is given by \cite{Ginzburg}

\begin{eqnarray}
f_{\gamma/e}(y)={{1}\over{g(\zeta)}}[1-y+{{1}\over{1-y}}
-{{4y}\over{\zeta(1-y)}}+{{4y^{2}}\over {\zeta^{2}(1-y)^{2}}}+
\lambda_{0}\lambda_{e} r\zeta (1-2r)(2-y)]
\end{eqnarray}
where

\begin{eqnarray}
g(\zeta)=&&g_{1}(\zeta)+
\lambda_{0}\lambda_{e}g_{2}(\zeta) \nonumber\\
g_{1}(\zeta)=&&(1-{{4}\over{\zeta}}
-{{8}\over{\zeta^{2}}})\ln{(\zeta+1)}
+{{1}\over{2}}+{{8}\over{\zeta}}-{{1}\over{2(\zeta+1)^{2}}} \\
g_{2}(\zeta)=&&(1+{{2}\over{\zeta}})\ln{(\zeta+1)}
-{{5}\over{2}}+{{1}\over{\zeta+1}}-{{1}\over{2(\zeta+1)^{2}}}
\end{eqnarray}
Here $r=y/[\zeta(1-y)]$ and $\zeta=4E_{e}E_{0}/M_{e}^{2}$. $E_{0}$
and $\lambda_{0}$ are the energy and the helicity of the initial
laser photon and $E_{e}$ and $\lambda_{e}$ are the energy and the
helicity of the initial electron beam before Compton backscattering.
$y$ is the fraction which represents the ratio between the scattered
photon and initial electron energy for the backscattered photons
moving along the initial electron direction. The maximum value of
$y$ reaches 0.83 when $\zeta=4.8$ in which the backscattered photon
energy is maximized without spoiling the luminosity. Backscattered
photons are not in fixed helicity states and their helicities are
given by a distribution:

\begin{eqnarray}
\xi(E_{\gamma},\lambda_{0})={{\lambda_{0}(1-2r)
(1-y+1/(1-y))+\lambda_{e} r\zeta[1+(1-y)(1-2r)^{2}]}
\over{1-y+1/(1-y)-4r(1-r)-\lambda_{e}\lambda_{0}r\zeta (2r-1)(2-y)}}
\end{eqnarray}
where $E_{\gamma}$ is the energy of backscattered photons. The
helicity-dependent differential cross section for the subprocess can
be written as

\begin{eqnarray}
d\hat \sigma(\lambda^{(1)}_0,\lambda^{(2)}_0
;\lambda_{W^+},\lambda_{W^-})
=\frac{1}{4}(1-\xi_1(E^{(1)}_{\gamma},\lambda^{(1)}_{0}))
(1-\xi_2(E^{(2)}_{\gamma},\lambda^{(2)}_{0}))d\hat
\sigma(-,-;\lambda_{W^+},\lambda_{W^-})\nonumber
\\+\frac{1}{4}(1-\xi_1(E^{(1)}_{\gamma},\lambda^{(1)}_{0}))
(1+\xi_2(E^{(2)}_{\gamma},\lambda^{(2)}_{0}))d\hat
\sigma(-,+;\lambda_{W^+},\lambda_{W^-})\nonumber
\\+\frac{1}{4}(1+\xi_1(E^{(1)}_{\gamma},\lambda^{(1)}_{0}))
(1-\xi_2(E^{(2)}_{\gamma},\lambda^{(2)}_{0}))d\hat
\sigma(+,-;\lambda_{W^+},\lambda_{W^-})\nonumber\\
+\frac{1}{4}(1+\xi_1(E^{(1)}_{\gamma},\lambda^{(1)}_{0}))
(1+\xi_2(E^{(2)}_{\gamma},\lambda^{(2)}_{0}))d\hat
\sigma(+,+;\lambda_{W^+},\lambda_{W^-})\nonumber
\\
\end{eqnarray}
Here
$d\hat{\sigma}(\lambda^{(1)}_{\gamma},\lambda^{(2)}_{\gamma};\lambda_{W^+},\lambda_{W^-})$
is the helicity-dependent differential cross section,
$\lambda^{(i)}_{\gamma}=+,-$ and $\lambda_{V}=+,-,0$
($V=W^+,\,W^-$). Superscripts (1) and (2) represent the incoming
gamma beams and $\xi_1(E^{(1)}_{\gamma},\lambda^{(1)}_{0})$ and
$\xi_2(E^{(2)}_{\gamma},\lambda^{(2)}_{0})$ represent the
corresponding helicity distributions. The integrated cross section
over the backscattered photon spectrums is given through the formula

\begin{eqnarray}
d\sigma(e^+ e^- \to \gamma \gamma \to
W^{+}W^{-}H)=\int_{z_{min}}^{z_{max}}dz \,2z \,
\int_{z^2/y_{max}}^{y_{max}}
\frac{dy}{y}f_{\gamma/e}(y)f_{\gamma/e}(z^2/y)\,d\hat{\sigma}(\gamma\gamma
\to W^{+}W^{-}H)\nonumber
\\
\end{eqnarray}
where, $d\hat{\sigma}(\gamma\gamma \to W^{+}W^{-}H)$ is the cross
section of the subprocess and the center of mass energy of the
$e^{+}e^{-}$ system, $\sqrt{s}$, is related to the center of mass
energy of the $\gamma\gamma$ system, $\sqrt{\hat{s}}$, by $\hat{s}
=z^2s$.

\begin{figure}
\includegraphics{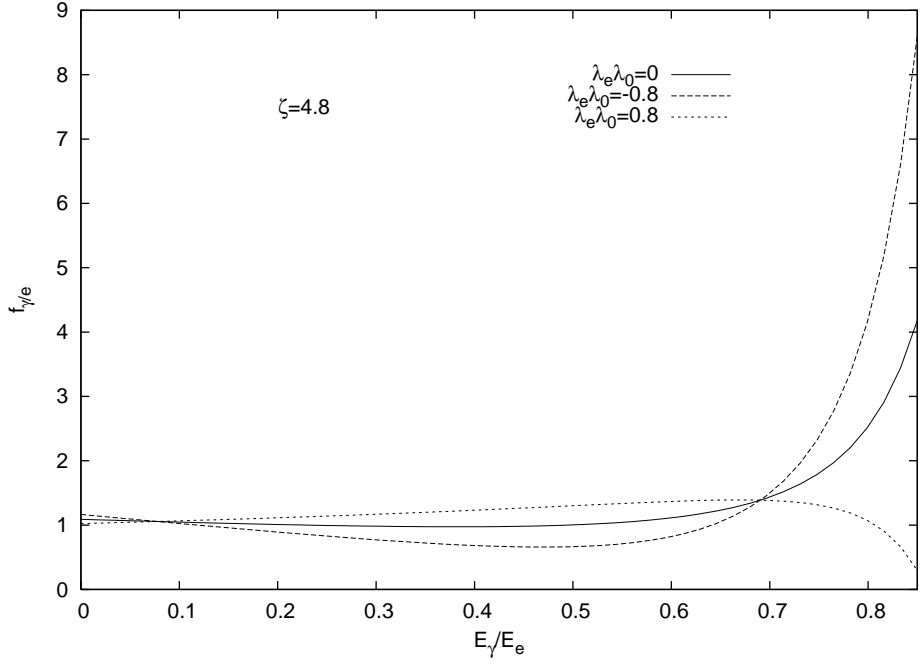}
\caption{Energy distribution of backscattered photons for
$\lambda_{e}\lambda_{0}=0,-0.8,0.8.$ \label{fig1}}
\end{figure}

In our calculations we accept that initial electron beam
polarizability is $|\lambda_{e}|=0.8$. To see the influence of
initial beam polarization, the energy distribution of backscattered
photons $f_{\gamma/e}$ is plotted for $\lambda_{e}\lambda_{0}=0,
-0.8$ and $+0.8$ in Fig.\ref{fig1}. It can be seen from the figure
that backscattered photon distribution is very low at high energies
in $\lambda_{e}\lambda_{0}=+0.8$. Therefore we will consider the
case $\lambda_{e}\lambda_{0}<0$ in the cross section calculations.
If we interchange backscattered photon helicities the cross section
does not change due to the symmetry. Moreover
$(\lambda^{(1)}_0,\lambda^{(1)}_e,\lambda^{(2)}_0,\lambda^{(2)}_e)=$
$(+1,-0.8,+1,-0.8)$ and
$(\lambda^{(1)}_0,\lambda^{(1)}_e,\lambda^{(2)}_0,\lambda^{(2)}_e)=$
$(-1,+0.8,-1,+0.8)$ combinations give the same cross section. So we
have two different combinations:
$(\lambda^{(1)}_0,\lambda^{(1)}_e,\lambda^{(2)}_0,\lambda^{(2)}_e)=$
$(+1,-0.8,+1,-0.8)$ and $(-1,+0.8,+1,-0.8)$.

\begin{figure}
\includegraphics{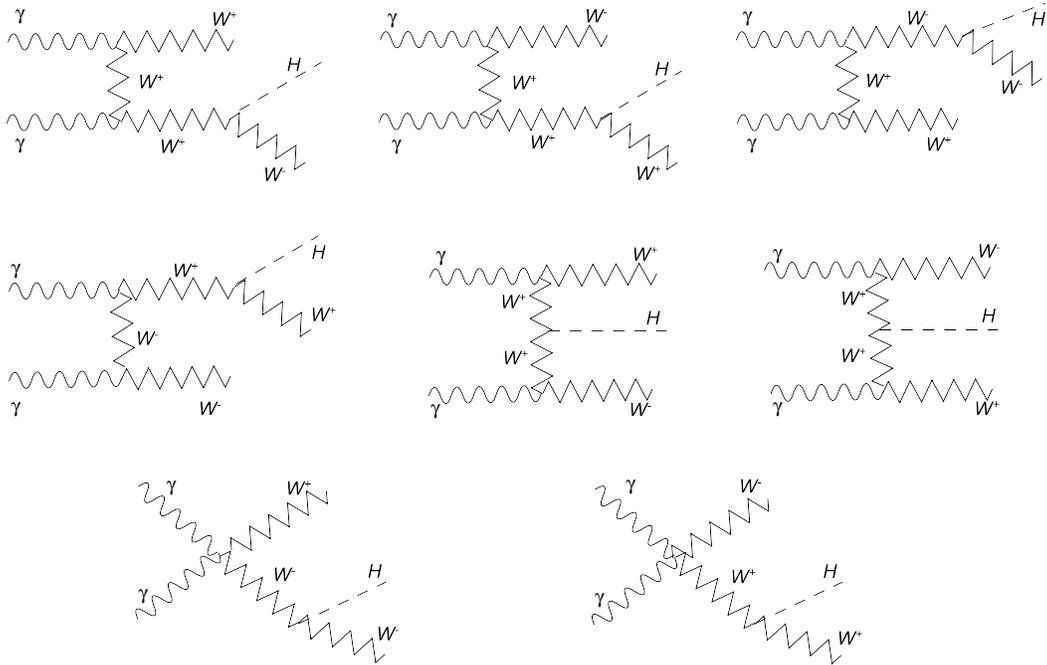}
\caption{Tree-level Feynman diagrams for $\gamma\gamma \to
W^{+}W^{-}H$ \label{fig2}}
\end{figure}

The process $\gamma\gamma \to W^{+}W^{-}H$ is described by eight
tree-level diagrams (Fig.\ref{fig2}). Each of the diagrams contains
an anomalous WWH vertex. Helicity amplitude techniques have been
used to obtain polarized amplitudes and the phase space integrations
have been performed by GRACE \cite{grace} which uses a Monte Carlo
routine.

\begin{figure}
\includegraphics{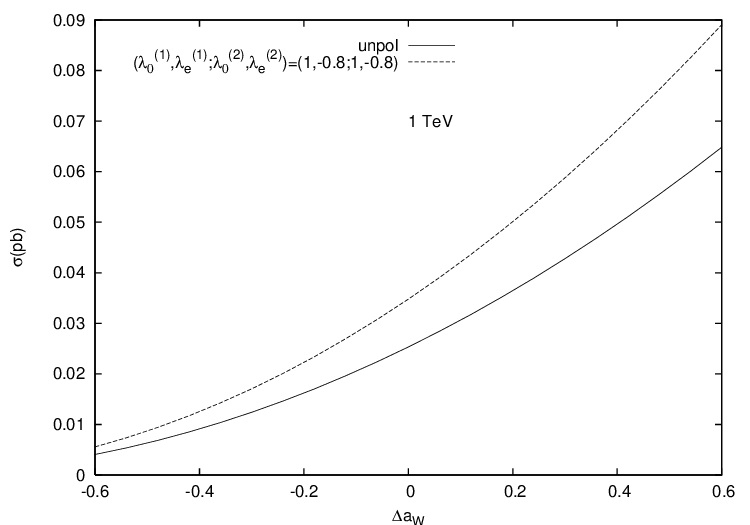}
\caption{The integrated total cross sections $\gamma\gamma \to
W^{+}W^{-}H$ as a function of anomalous coupling $\Delta a_{W}$ for
$\sqrt{s}=1 TeV$. The legends are for initial beam polarizations.
\label{fig3}}
\end{figure}

\begin{figure}
\includegraphics{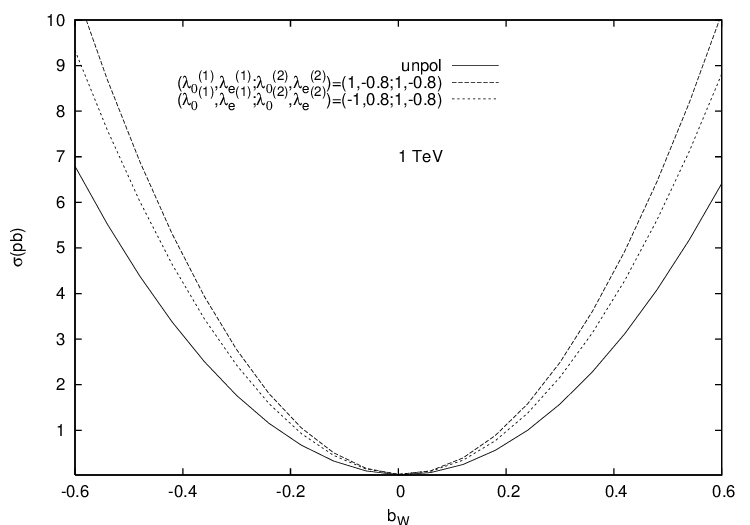}
\caption{The integrated total cross sections $\gamma\gamma \to
W^{+}W^{-}H$ as a function of anomalous coupling $b_{W}$ for
$\sqrt{s}=1 TeV$. The legends are for initial beam polarizations.
\label{fig4}}
\end{figure}

\begin{figure}
\includegraphics{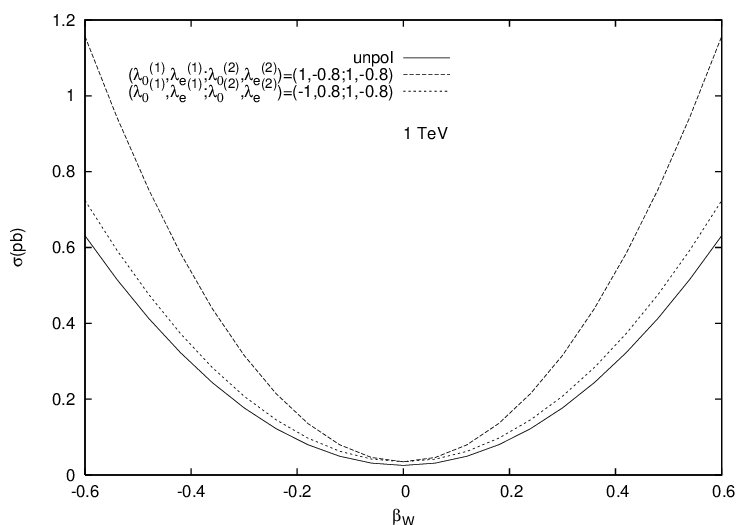}
\caption{The integrated total cross sections $\gamma\gamma \to
W^{+}W^{-}H$ as a function of anomalous coupling $\beta_{W}$ for
$\sqrt{s}=1 TeV$. The legends are for initial beam polarizations.
\label{fig5}}
\end{figure}

\begin{figure}
\includegraphics{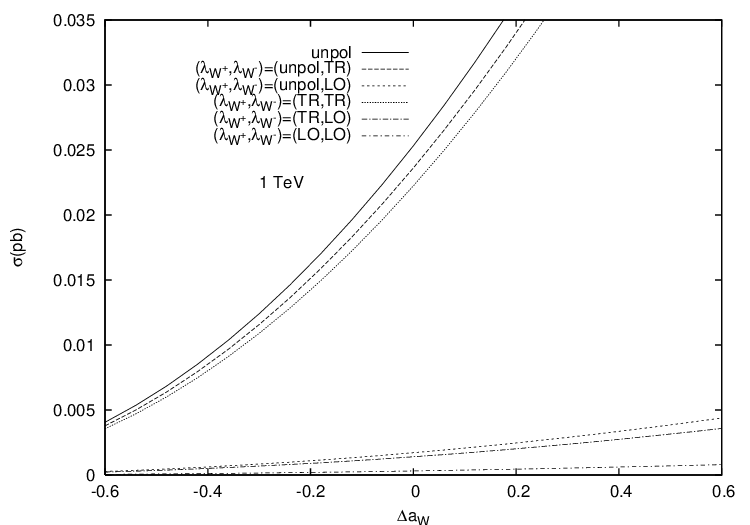}
\caption{The integrated total cross sections $\gamma\gamma \to
W^{+}W^{-}H$ as a function of anomalous coupling $\Delta a_{W}$ for
$\sqrt{s}=1 TeV$. The legends are for final state polarizations.
\label{fig6}}
\end{figure}

\begin{figure}
\includegraphics{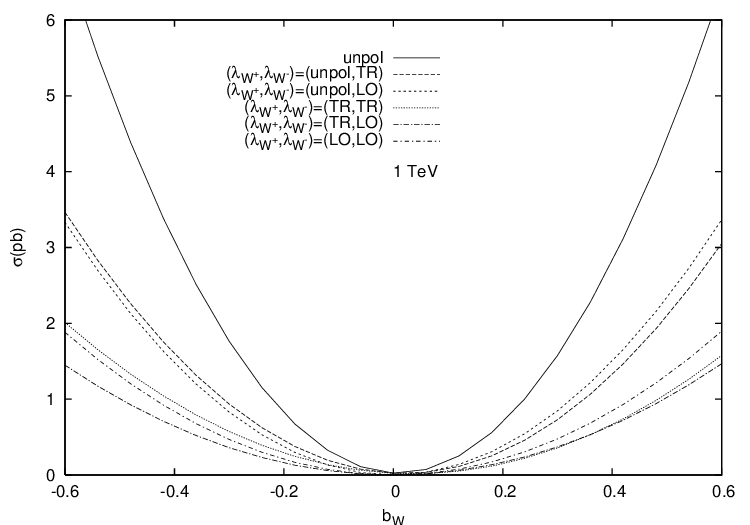}
\caption{The integrated total cross sections $\gamma\gamma \to
W^{+}W^{-}H$ as a function of anomalous coupling $b_{W}$ for
$\sqrt{s}=1 TeV$. The legends are for final state polarizations.
\label{fig7}}
\end{figure}

\begin{figure}
\includegraphics{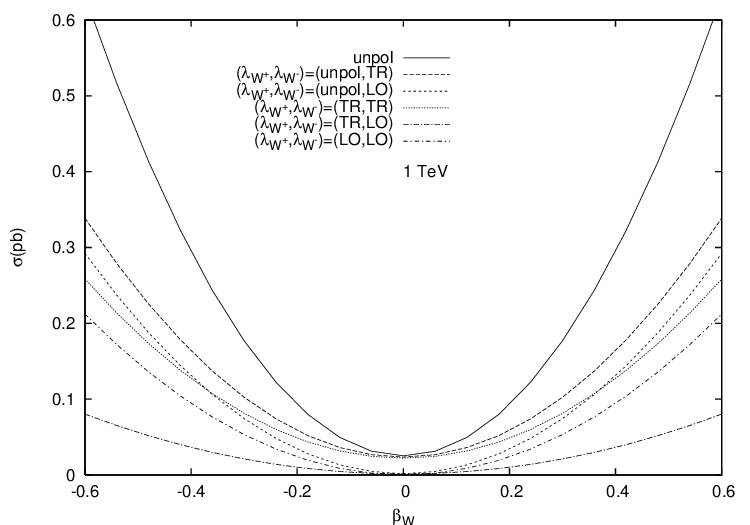}
\caption{The integrated total cross sections $\gamma\gamma \to
W^{+}W^{-}H$ as a function of anomalous coupling $\beta_{W}$ for
$\sqrt{s}=1 TeV$. The legends are for final state polarizations.
\label{fig8}}
\end{figure}

One can see from Figs.\ref{fig3}-\ref{fig5} the influence of the
initial state polarizations on the deviations of the total cross
sections from their SM value at $\sqrt{s}=1$TeV. In Fig.\ref{fig3}
the initial state polarization configuration
$(\lambda^{(1)}_0,\lambda^{(1)}_e,\lambda^{(2)}_0,\lambda^{(2)}_e)=(+1,-0.8,+1,-0.8)$
coincides with $(-1,+0.8,+1,-0.8)$, therefore we plot one of them.
We see from Fig.\ref{fig4} and Fig.\ref{fig5} that cross section for
$(\lambda^{(1)}_0,\lambda^{(1)}_e,\lambda^{(2)}_0,\lambda^{(2)}_e)=(+1,-0.8,+1,-0.8)$
 is more sensitive to the anomalous coupling parameters
$b_{W}$ and $\beta_{W}$ than $(-1,+0.8,+1,-0.8)$. In
Figs.\ref{fig6}-\ref{fig8} we plot the total cross section as a
function of anomalous parameters for various final state
polarizations. In these figures TR and LO stand for "transverse" and
"longitudinal" respectively. We consider all possible polarization
combinations for the final $W^{+}$ and $W^{-}$ bosons. In
Fig.\ref{fig6} cross sections are plotted as a function of the
parameter $\Delta a_{W}$. It is clear from the figure that the
unpolarized cross section drastically grows as the parameter $\Delta
a_{W}$ increases. On the other hand, this growth in the cross
section is relatively small for polarized cases. Therefore from
Fig.\ref{fig6} we see that polarized cross sections are insensitive
to the parameter $\Delta a_{W}$.

It can be extracted from figures that the most sensitive
polarization configurations are
$(\lambda_{W^{+}},\lambda_{W^{-}})=$(LO,LO) and (TR,LO). For
instance in Fig.\ref{fig7}, SM and anomalous cross sections at
$b_{W}=0.12$ are $\sigma_{SM}=3.12\times10^{-4}$pb and
$\sigma(b_{W}=0.12)=7.73\times10^{-2}$pb respectively for the
polarization configuration (LO,LO). For the (TR,LO) case they are
$\sigma_{SM}=1.39\times10^{-3}$pb and
$\sigma(b_{W}=0.12)=6.16\times10^{-2}$pb. Therefore we see that
cross sections at the polarization configuration (LO,LO) and (TR,LO)
increase by factor of 248 and 44 as $b_{W}$ increases from 0 to
0.12. But this increment is only a factor of 10 in the unpolarized
case; $\sigma_{SM}=2.5\times10^{-2}$pb and
$\sigma(b_{W}=0.12)=2.5\times10^{-1}$pb. It can be seen from
Fig.\ref{fig8} that cross sections have a symmetric behavior as a
function of the anomalous parameter $\beta_{W}$. Longitudinal W
boson polarizations improve the deviations from the SM at both
positive and negative values of $\beta_{W}$.

\section{Angular correlations for final state fermions}

The angular distributions of $W^{+}$ and $W^{-}$ decay products have
clear correlations with the helicity states of these final state
gauge bosons. Therefore, in principle, polarization states of final
$W^{+}$ and $W^{-}$ bosons can be determined by measuring the
angular distributions of $W^{+}$ and $W^{-}$ decay products. This
kind of analysis was done in reference \cite{hagiwara2} for the
process $e^{+}e^{-}\to W^{+}W^{-}$. We consider the differential
cross section for the complete process,

\begin{eqnarray}
\gamma(k_{1},\lambda_{\gamma}^{(1)})+\gamma(k_{2},\lambda_{\gamma}^{(2)})
\to&&
W^{-}(q_{1},\lambda_{W^{-}})+W^{+}(q_{2},\lambda_{W^{+}})+H(q_{3})
\nonumber\\
&&W^{-}(q_{1},\lambda_{W^{-}})\to f_{1}(p_{1},\sigma_{1})
\bar{f}_{2}(p_{2},\sigma_{2})
\nonumber\\
&&W^{+}(q_{2},\lambda_{W^{+}})\to f_{3}(p_{3},\sigma_{3})
\bar{f}_{4}(p_{4},\sigma_{4})
\end{eqnarray}
with massless fermions $f_{1},\bar{f}_{2},f_{3},\bar{f}_{4}$. Here
$\lambda_{\gamma}^{(1)}$ and $\lambda_{\gamma}^{(2)}$ are the
incoming photon helicities, $\lambda_{W^{-}}$ and $\lambda_{W^{+}}$
are the outgoing $W^{-}$ and $W^{+}$ helicities. $\sigma_{i}$
represent the helicities of final fermions $f_{i}$ or $\bar{f_{i}}$.

The full amplitude can be written as follows:

\begin{eqnarray}
M(k_{1},\lambda_{\gamma}^{(1)};k_{2},\lambda_{\gamma}^{(2)};q_{3};p_{i},\sigma_{i})=&&
D_{W^{-}}(q_{1}^{2})D_{W^{+}}(q_{2}^{2})\sum_{\lambda_{W^{-}}}\sum_{\lambda_{W^{+}}}
M_{1}(k_{1},\lambda_{\gamma}^{(1)};k_{2},\lambda_{\gamma}^{(2)};q_{1},\lambda_{W^{-}};
q_{2},\lambda_{W^{+}};q_{3})
\nonumber\\
&&\times
M_{2}(q_{1},\lambda_{W^{-}};p_{1},\sigma_{1};p_{2},\sigma_{2})\times
M_{3}(q_{2},\lambda_{W^{+}};p_{3},\sigma_{3};p_{4},\sigma_{4})
\end{eqnarray}
where
$M_{1}(k_{1},\lambda_{\gamma}^{(1)};k_{2},\lambda_{\gamma}^{(2)};q_{1},\lambda_{W^{-}};
q_{2},\lambda_{W^{+}};q_{3})$ is the production amplitude for
$\gamma\gamma \to W^{+}W^{-}H$ with on-shell $W^{-}$ and $W^{+}$.
$M_{2}(q_{1},\lambda_{W^{-}};p_{1},\sigma_{1};p_{2},\sigma_{2})$ and
$M_{3}(q_{2},\lambda_{W^{+}};p_{3},\sigma_{3};p_{4},\sigma_{4})$ are
the decay amplitudes of $W^{-}$ and $W^{+}$ to fermions.
$D_{W^{-}}(q_{1}^{2})$ and $D_{W^{+}}(q_{2}^{2})$ are the
Breit-Wigner propagator factors for $W^{-}$ and $W^{+}$ bosons.

In this work we consider the hadronic decay channel of final state
bosons. Therefore $f_{1},\bar{f}_{2},f_{3},\bar{f}_{4}$ are quarks.
$M_{2}$ and $M_{3}$ decay amplitudes are expressed in the rest frame
of $W^{-}$ and $W^{+}$ respectively. In the $W^{-}$ rest frame, we
parametrize $f_{1}$ and $\bar{f}_{2}$ four-momenta as

\begin{eqnarray}
p_{1}^{\mu}=&&\frac{m_{W}}{2}(1,sin\theta cos\phi,sin\theta
sin\phi,cos\theta)
\nonumber\\
p_{2}^{\mu}=&&\frac{m_{W}}{2}(1,-sin\theta cos\phi,-sin\theta
sin\phi,-cos\theta)
\end{eqnarray}

and in the $W^{+}$ rest frame, we choose the antiparticle
($\bar{f}_{4}$) angles as $\bar{\theta}$ and $\bar{\phi}$,

\begin{eqnarray}
p_{3}^{\mu}=&&\frac{m_{W}}{2}(1,-sin\bar{\theta}
cos\bar{\phi},-sin\bar{\theta} sin\bar{\phi},-cos\bar{\theta})
\nonumber\\
p_{4}^{\mu}=&&\frac{m_{W}}{2}(1,sin\bar{\theta}
cos\bar{\phi},sin\bar{\theta} sin\bar{\phi},cos\bar{\theta})
\end{eqnarray}

In this convention the angles of the d-type quark are chosen as
$(\theta,\phi)$ in $W^{-}$ decays and $(\bar{\theta},\bar{\phi})$ in
$W^{+}$ decays. $M_{2}$  and $M_{3}$ decay amplitudes are given by

\begin{eqnarray}
M_{2}=g_{e}g_{-}^{Wf_{1}f_{2}}Cm_{W}\delta_{\sigma_{1,-}}\delta_{\sigma_{2,+}}l_{\lambda_{W^{-}}}
\end{eqnarray}
\begin{eqnarray}
M_{3}=-g_{e}g_{-}^{Wf_{3}f_{4}}\bar{C}m_{W}\delta_{\sigma_{3,-}}\delta_{\sigma_{4,+}}\bar{l}_{\lambda_{W^{+}}}
\end{eqnarray}
where
\begin{eqnarray}
(l_{-},l_{0},l_{+})=(\frac{1}{\sqrt{2}}(1+cos\theta)e^{-i\phi},-sin\theta,\frac{1}{\sqrt{2}}(1-cos\theta)e^{i\phi})
\end{eqnarray}
\begin{eqnarray}
(\bar{l}_{-},\bar{l}_{0},\bar{l}_{+})=(\frac{1}{\sqrt{2}}(1+cos\bar{\theta})e^{i\bar{\phi}},-sin\bar{\theta},\frac{1}{\sqrt{2}}(1-cos\bar{\theta})e^{-i\bar{\phi}})
\end{eqnarray}
Here $g_{-}^{Wf_{1}f_{2}}$ and $g_{-}^{Wf_{3}f_{4}}$ are the
standard V-A coupling for quarks
($g_{-}^{Wf_{1}f_{2}}=g_{-}^{Wf_{3}f_{4}}=U_{ij}/\sqrt{2}sin\theta_{W}$).
$C$ and $\bar{C}$ denote the effective color factors ($\sqrt{3}$)
for hadronic decay processes of the $W^{-}$ and $W^{+}$.

Polarization summed squared matrix elements are given by

\begin{eqnarray}
\sum_{\lambda_{\gamma}^{(1)},\lambda_{\gamma}^{(2)},\sigma_{i}}
|M(k_{1},\lambda_{\gamma}^{(1)};k_{2},\lambda_{\gamma}^{(2)};q_{3};p_{i},\sigma_{i})|^{2}
=|D_{W^{-}}(q_{1}^{2})|^{2}|D_{W^{+}}(q_{2}^{2})|^{2}
P_{\lambda'_{W^{-}}\lambda'_{W^{+}}}^{\lambda_{W^{-}}\lambda_{W^{+}}}
D_{\lambda'_{W^{-}}}^{\lambda_{W^{-}}}\bar{D}_{\lambda'_{W^{+}}}^{\lambda_{W^{+}}}
\end{eqnarray}

In this equation summation over repeated indices
$(\lambda_{W^{-}},\lambda'_{W^{-}},\lambda_{W^{+}},\lambda'_{W^{+}})=+,-,0$
is implied.
$P_{\lambda'_{W^{-}}\lambda'_{W^{+}}}^{\lambda_{W^{-}}\lambda_{W^{+}}}$
is the production tensor and
$D_{\lambda'_{W^{-}}}^{\lambda_{W^{-}}}$,
$\bar{D}_{\lambda'_{W^{+}}}^{\lambda_{W^{+}}}$ are the decay tensors
for $W^{-}$ and $W^{+}$ boson respectively. They are defined by

\begin{eqnarray}
P_{\lambda'_{W^{-}}\lambda'_{W^{+}}}^{\lambda_{W^{-}}\lambda_{W^{+}}}=&&
\sum_{\lambda_{\gamma}^{(1)},\lambda_{\gamma}^{(2)}}M_{1}
(k_{1},\lambda_{\gamma}^{(1)};k_{2},\lambda_{\gamma}^{(2)};q_{1},\lambda_{W^{-}};q_{2},\lambda_{W^{+}};q_{3})
\nonumber\\
&&\times M_{1}^{*}
(k_{1},\lambda_{\gamma}^{(1)};k_{2},\lambda_{\gamma}^{(2)};q_{1},\lambda'_{W^{-}};q_{2},\lambda'_{W^{+}};q_{3})
\end{eqnarray}
\begin{eqnarray}
D_{\lambda'_{W^{-}}}^{\lambda_{W^{-}}}=\sum_{\sigma_{1},\sigma_{2}}M_{2}(q_{1},\lambda_{W^{-}};p_{1},\sigma_{1};
p_{2},\sigma_{2})M^{*}_{2}(q_{1},\lambda'_{W^{-}};p_{1},\sigma_{1};
p_{2},\sigma_{2})
\end{eqnarray}
\begin{eqnarray}
\bar{D}_{\lambda'_{W^{+}}}^{\lambda_{W^{+}}}=\sum_{\sigma_{3},\sigma_{4}}M_{3}(q_{2},\lambda_{W^{+}};p_{3},\sigma_{3};
p_{4},\sigma_{4})M^{*}_{3}(q_{2},\lambda'_{W^{+}};p_{3},\sigma_{3};
p_{4},\sigma_{4})
\end{eqnarray}
The differential cross section can be written in the following form:

\begin{eqnarray}
d\sigma=&&\frac{1}{2s}|M|^{2}\frac{d^{3}p_{1}}{(2\pi)^{3}2E_{p_{1}}}\frac{d^{3}p_{2}}{(2\pi)^{3}2E_{p_{2}}}
\frac{d^{3}p_{3}}{(2\pi)^{3}2E_{p_{3}}}\frac{d^{3}p_{4}}{(2\pi)^{3}2E_{p_{4}}}\frac{d^{3}q_{3}}{(2\pi)^{3}2E_{q_{3}}}
\nonumber\\
&&\times(2\pi)^{4}\delta^{4}(k_{1}+k_{2}-p_{1}-p_{2}-p_{3}-p_{4}-q_{3})
\end{eqnarray}

Using narrow width approximation it is straightforward to express
the differential cross section as

\begin{eqnarray}
d\sigma=&&\frac{1}{2s}(2\pi)^{4}\delta^{4}(k_{1}+k_{2}-q_{1}-q_{2}-q_{3})\frac{\pi^{2}}{2^{6}(2\pi)^{6}\Gamma_{W^{-}}\Gamma_{W^{+}}m_{W}^{2}}
\nonumber\\
&&\times
P_{\lambda'_{W^{-}}\lambda'_{W^{+}}}^{\lambda_{W^{-}}\lambda_{W^{+}}}D_{\lambda'_{W^{-}}}^{\lambda_{W^{-}}}
\bar{D}_{\lambda'_{W^{+}}}^{\lambda_{W^{+}}}\frac{d^{3}q_{1}}{(2\pi)^{3}2E_{q_{1}}}\frac{d^{3}q_{2}}{(2\pi)^{3}2E_{q_{2}}}
\frac{d^{3}q_{3}}{(2\pi)^{3}2E_{q_{3}}}dcos\theta d\phi
dcos\bar{\theta} d\bar{\phi}
\end{eqnarray}

After integration over azimuthal angles $\phi$ and $\bar{\phi}$
interference terms will vanish and only the diagonal terms
$\lambda_{W^{-}}=\lambda'_{W^{-}}$ and
$\lambda_{W^{+}}=\lambda'_{W^{+}}$ will survive. The differential
cross section can be written as

\begin{eqnarray}
d\sigma=d\sigma_{1}(\lambda_{W^{-}},\lambda_{W^{+}})d_{\lambda_{W^{-}}}^{\lambda_{W^{-}}}\bar{d}_{\lambda_{W^{+}}}^{\lambda_{W^{+}}}
\frac{9}{16}B(W^{-}\to f_{1}\bar{f}_{2})B(W^{+}\to
f_{3}\bar{f}_{4})dcos\theta dcos\bar{\theta}
\end{eqnarray}

Here $d\sigma_{1}(\lambda_{W^{-}},\lambda_{W^{+}})$ is the helicity-
dependent production cross section, $B(W^{-}\to f_{1}\bar{f}_{2})$
and $B(W^{+}\to f_{3}\bar{f}_{4})$ are the branching ratios of W
bosons to quarks. $d_{\lambda_{W^{-}}}^{\lambda_{W^{-}}}$ and
$\bar{d}_{\lambda_{W^{+}}}^{\lambda_{W^{+}}}$ are related to the
diagonal elements of decay tensors (19-20) as

\begin{eqnarray}
d_{\lambda_{W^{-}}}^{\lambda_{W^{-}}}=l_{\lambda_{W^{-}}}l^{*}_{\lambda_{W^{-}}}
\nonumber\\
\bar{d}_{\lambda_{W^{+}}}^{\lambda_{W^{+}}}=\bar{l}_{\lambda_{W^{+}}}\bar{l}^{*}_{\lambda_{W^{+}}}
\end{eqnarray}

The production cross section has nine different polarization
configurations and identification of all nine polarized cross
sections is difficult because of the necessity of charge (flavor)
identification of both the $W^{-}$ and $W^{+}$ decay products.
Experimentally, in view of the difficulty of flavor identification
there is an ambiguity in reconstruction of the polar angles of decay
products. This makes it very difficult to identify polarization
states $\lambda_{W^+}=+1,-1$ and $\lambda_{W^-}=+1,-1$ separately.
On the other hand, cross section for the transverse polarization
state which is the sum of $\sigma(\lambda_{W}=+1)$ and
$\sigma(\lambda_{W}=-1)$ can be determined without an ambiguity.
This is also true for the longitudinal polarization case. Therefore
it is reasonable to claim that longitudinal (LO) and transverse (TR)
polarizations can be identified \cite{hagiwara2}. Thus we define the
following cross sections:

\begin{eqnarray}
d\sigma_{1}(TR,TR)=&&\sum_{\lambda_{W^{-}}=+,-}\sum_{\lambda_{W^{+}}=+,-}d\sigma_{1}(\lambda_{W^{-}},\lambda_{W^{+}})\\
d\sigma_{1}(LO,LO)=&&d\sigma_{1}(0,0)\\
d\sigma_{1}(TR,LO)=&&\sum_{\lambda_{W^{-}}=+,-}d\sigma_{1}(\lambda_{W^{-}},0)\\
d\sigma_{1}(LO,TR)=&&\sum_{\lambda_{W^{+}}=+,-}d\sigma_{1}(0,\lambda_{W^{+}})\\
d\sigma_{1}(TR,unpol)=&&d\sigma_{1}(TR,TR)+d\sigma_{1}(TR,LO)\\
d\sigma_{1}(LO,unpol)=&&d\sigma_{1}(LO,TR)+d\sigma_{1}(LO,LO)\\
d\sigma_{1}(unpol,TR)=&&d\sigma_{1}(TR,TR)+d\sigma_{1}(LO,TR)\\
d\sigma_{1}(unpol,LO)=&&d\sigma_{1}(TR,LO)+d\sigma_{1}(LO,LO)
\end{eqnarray}

\begin{figure}
\includegraphics{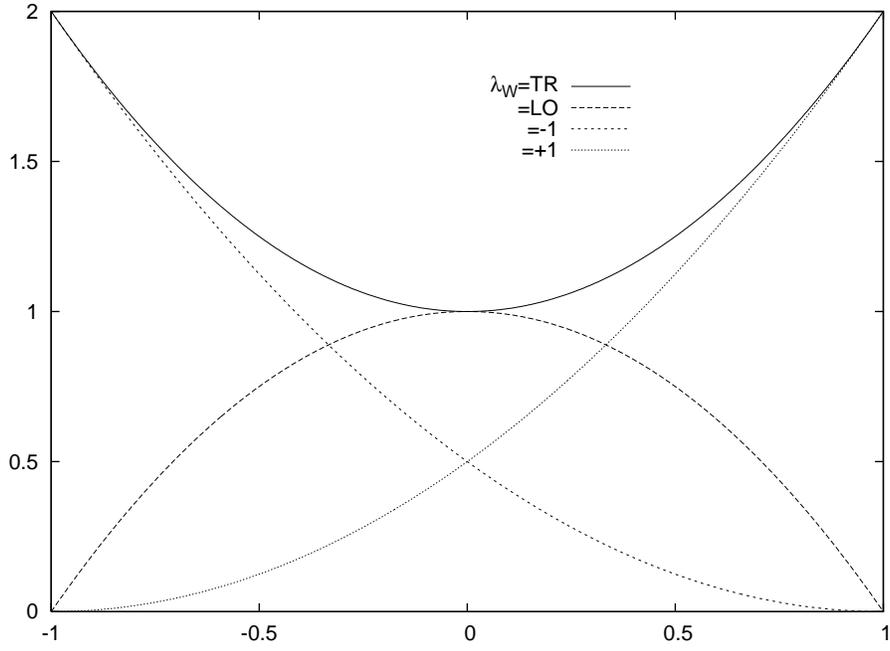}
\caption{$d_{\lambda_{W^{-}}}^{\lambda_{W^{-}}}$ versus $cos\theta$.
The legends are for various polarization states of the final $W^{-}$
boson. \label{fig9}}
\end{figure}

For fixed $W^{-}$ and $W^{+}$ helicities above cross sections can be
obtained from a fit to polar angle distributions of the $W^{-}$ and
$W^{+}$ decay products in the $W^{-}$ and $W^{+}$ rest frames. More
specifically, for $\lambda_{W^{-}}=\pm1,0$ polarization states of
final $W^{-}$, production cross sections
$d\sigma_{1}(\pm,\lambda_{W^{+}})$ and
$d\sigma_{1}(0,\lambda_{W^{+}})$ can be obtained from a fit to
$d_{+}^{+}$, $d_{-}^{-}$ and $d_{0}^{0}$ distributions in the
$W^{-}$ rest frame (eqn.(23)). Similarly production cross sections
$d\sigma_{1}(\lambda_{W^{-}},\pm)$ and
$d\sigma_{1}(\lambda_{W^{-}},0)$ can be obtained from a fit to
$\bar{d}_{+}^{+}$, $\bar{d}_{-}^{-}$ and $\bar{d}_{0}^{0}$
distributions in the $W^{+}$ rest frame. In Fig. \ref{fig9}
$d_{\lambda_{W^{-}}}^{\lambda_{W^{-}}}$ distribution is plotted for
various polarization states of final $W^{-}$ boson. As can be seen
from the figure, longitudinal (LO) and transverse (TR) distributions
are well separated from each other.

\section{Sensitivity to anomalous couplings}

We have obtained 95\% C.L. limits on the anomalous coupling
parameters $\Delta a_{W}$, $b_{W}$ and $\beta_{W}$ using a simple
$\chi^{2}$ analysis at $\sqrt{s}$ = 0.5 and 1 TeV energies and an
integrated luminosity $L_{int}=500 fb^{-1}$ without systematic
errors. All our numerical calculations are for a Higgs mass of 120
GeV, hence the dominant decay mode should be $H \to b\bar{b}$ with a
branching ratio $B_{H}\approx0.9$.

In the literature there have been several experimental studies for
the measurement of W polarization \cite{abbiendi}. Angular
distribution of the W boson decay products has a clear correlation
with the helicity states of it. Therefore it is reasonable to assume
that W boson polarization can be measured. We consider the case in
which W momentum is reconstructible. We take into account the $W \to
q\bar{q'}$ decay channel with a branching ratio $B_{W}\approx0.68$.
The expected number of events are given by
$N=E(B_{W})^{2}B_{H}L_{int}\sigma$, where E is the b-tagging
efficiency and it is taken to be 0.7 as in Refs.
\cite{biswal,choudhury}.

In Table \ref{tab1}-\ref{tab2} we show $95\%$ C.L. sensitivity
limits on the anomalous coupling parameters $\Delta a_{W}$, $b_{W}$
and $\beta_{W}$ for $\sqrt{s}=1$ and 0.5 TeV energies. In the
tables, LO, TR represent the longitudinal, transverse polarization
states and TR+LO describes the unpolarized $W^{+}$ and $W^{-}$
bosons,
$(\lambda^{(1)}_0,\lambda^{(1)}_e,\lambda^{(2)}_0,\lambda^{(2)}_e)=(0,0,0,0)$
stand for the unpolarized initial beams.

We see from Table \ref{tab1} that polarization leads to a
significant improvement on the upper bound of $b_{W}$. Polarized
bounds are compared with unpolarized bounds which are given in the
first line of the table. The initial state
$(\lambda^{(1)}_0,\lambda^{(1)}_e,\lambda^{(2)}_0,\lambda^{(2)}_e)=(-1,+0.8,+1,-0.8)$
together with the final state
$(\lambda_{W^{+}},\lambda_{W^{-}})=$(LO,LO) polarization
configuration improves the upper bound of $b_{W}$ by a factor of
7.8. Final state (TR,TR) polarization combined with initial state
polarizations improves the lower bound of $b_{W}$ by a factor of
1.3. $(+1,-0.8,+1,-0.8)$ initial state with (LO, LO) final state
polarization configuration improves both upper and lower bounds of
$\beta_{W}$ by a factor of 2.8 at $\sqrt{s}=1$ TeV. These
improvement factors are smaller for $\sqrt{s}=0.5$ TeV. In Table
\ref{tab2}, $(-1,+0.8,+1,-0.8)$ together with (LO,TR+LO) improves
the upper bound of $b_{W}$ by a factor of 2.3. However limits on
$\Delta a_{W}$ are improved slightly by the initial state
polarizations.

As we have mentioned in the introduction, WWH couplings are isolated
also by the process $e^{-}\gamma\to \nu_{e}W^{-}H$ \cite{choudhury}.
Our limits on the upper bound of $b_{W}$ are a factor from 6 to 8.7
better than the bounds obtained in $e^{-}\gamma\to \nu_{e}W^{-}H$
depending on energy. The lower bound of $b_{W}$ is approximately 1.7
times better at $\sqrt{s}=1$ TeV. On the other hand at $\sqrt{s}=1$
TeV, our bounds on $\beta_{W}$ are 1.8 times better than the bounds
acquired in $e^{-}\gamma\to \nu_{e}W^{-}H$.

For further analysis one can consider the systematic errors. The
expected sources of systematic errors may result from the
uncertainty on the measurement of $\gamma\gamma$ luminosity,
helicity of incoming photons after Compton backscattering and
uncertainty on the photon spectra. Moreover, the systematic
uncertainties on the measurement of angular distributions of the
decay products of the W bosons can be considered. For more precise
results, further analysis needs to be supplemented with a more
detailed knowledge of the experimental conditions.

In conclusion, we have obtained a considerable improvement in the
sensitivity bounds of the anomalous parameters $b_{W}$ and
$\beta_{W}$ by taking into account incoming beam polarizations and
the final state polarizations of the W bosons. The subprocess
$\gamma\gamma \to W^{+}W^{-}H$ in the $\gamma\gamma$ mode of a
linear collider isolates WWH couplings and provide sensitive limits.
Thus $\gamma\gamma$ colliders are better equipped than $e^{+}e^{-}$
and $e\gamma$ colliders to study these couplings.

\begin{acknowledgments}
The author acknowledges support through The Scientific and
Technological Research Council of Turkey (TUBITAK) BIDEB-2219 grant.
\end{acknowledgments}


\pagebreak

\begin{table}
\caption{Sensitivity of the $\gamma\gamma$ collision to anomalous
WWH couplings at 95\% C.L. for $\sqrt{s}=1$ TeV and $L_{int}=500$
fb$^{-1}$. The effects of initial beam polarizations and final state
$W^{+}$, $W^{-}$ polarizations are shown in each row. Only one of
the couplings is assumed to deviate from the SM at a time.
\label{tab1}}
\begin{ruledtabular}
\begin{tabular}{ccccccccc}
$\lambda_{0}^{(1)}$&$\lambda_{e}^{(1)}$&$\lambda_{0}^{(2)}$&$\lambda_{e}^{(2)}$&$\lambda_{W^{+}}$&
$\lambda_{W^{-}}$&$\Delta a_{W}$&$b_{W}$&$\beta_{W}$ \\
\hline

0&0&0&0&TR+LO&TR+LO&(-0.0164, 0.0162)&(-0.0023, 0.0194)&(-0.0220, 0.0220)\\
\hline
0&0&0&0&TR&TR+LO&(-0.0170, 0.0167)&(-0.0022, 0.0398)&(-0.0300, 0.0300)\\
0&0&0&0&LO&TR+LO&(-0.0646, 0.0607)&(-0.0065, 0.0035)&(-0.0163, 0.0163)\\
0&0&0&0&TR&TR&(-0.0175, 0.0172)&(-0.0021, 0.0743)&(-0.0340, 0.0340)\\
0&0&0&0&TR&LO&(-0.0717, 0.0669)&(-0.0093, 0.0051)&(-0.0295, 0.0295)\\
0&0&0&0&LO&LO&(-0.1589, 0.1369)&(-0.0055, 0.0032)&(-0.0124, 0.0124)\\
+1&-0.8&+1&-0.8&TR+LO&TR+LO&(-0.0140, 0.0138)&(-0.0019, 0.0179)&(-0.0176, 0.0176)\\
-1&+0.8&+1&-0.8&TR+LO&TR+LO&(-0.0140, 0.0138)&(-0.0020, 0.0190)&(-0.0225, 0.0225)\\
+1&-0.8&+1&-0.8&TR&TR+LO&(-0.0144, 0.0142)&(-0.0019, 0.0399)&(-0.0270, 0.0270)\\
+1&-0.8&+1&-0.8&LO&TR+LO&(-0.0595, 0.0561)&(-0.0043, 0.0033)&(-0.0113, 0.0113)\\
-1&+0.8&+1&-0.8&TR&TR+LO&(-0.0144, 0.0142)&(-0.0019, 0.0384)&(-0.0282, 0.0282)\\
-1&+0.8&+1&-0.8&LO&TR+LO&(-0.0569, 0.0539)&(-0.0070, 0.0028)&(-0.0183, 0.0183)\\
+1&-0.8&+1&-0.8&TR&LO&(-0.0652, 0.0612)&(-0.0072, 0.0050)&(-0.0250, 0.0250)\\
-1&+0.8&+1&-0.8&TR&LO&(-0.0627, 0.0590)&(-0.0086, 0.0041)&(-0.0271, 0.0271)\\
+1&-0.8&+1&-0.8&TR&TR&(-0.0148, 0.0145)&(-0.0018, 0.0759)&(-0.0318, 0.0318)\\
+1&-0.8&+1&-0.8&LO&LO&(-0.1571, 0.1356)&(-0.0032, 0.0028)&(-0.0079, 0.0079)\\
-1&+0.8&+1&-0.8&TR&TR&(-0.0148, 0.0146)&(-0.0018, 0.0770)&(-0.0322, 0.0322)\\
-1&+0.8&+1&-0.8&LO&LO&(-0.1419, 0.1241)&(-0.0066, 0.0025)&(-0.0155, 0.0155)\\

\end{tabular}
\end{ruledtabular}
\end{table}

\begin{table}
\caption{The same as Table I but for $\sqrt{s}=0.5$ TeV.
\label{tab2}}
\begin{ruledtabular}
\begin{tabular}{ccccccccc}
$\lambda_{0}^{(1)}$&$\lambda_{e}^{(1)}$&$\lambda_{0}^{(2)}$&$\lambda_{e}^{(2)}$&$\lambda_{W^{+}}$&
$\lambda_{W^{-}}$&$\Delta a_{W}$&$b_{W}$&$\beta_{W}$ \\
\hline

0&0&0&0&TR+LO&TR+LO&(-0.1239, 0.1102)&(-0.0207, 0.0161)&(-0.0725, 0.0725)\\
\hline
0&0&0&0&TR&TR+LO&(-0.1424, 0.1245)&(-0.0182, 0.0246)&(-0.0813, 0.0813)\\
0&0&0&0&LO&TR+LO&(-0.2844, 0.2198)&(-0.0373, 0.0122)&(-0.0912, 0.0912)\\
0&0&0&0&TR&TR&(-0.1605, 0.1381)&(-0.0159, 0.0370)&(-0.0888, 0.0888)\\
0&0&0&0&TR&LO&(-0.3635, 0.2629)&(-0.0411, 0.0153)&(-0.1082, 0.1082)\\
0&0&0&0&LO&LO&(-0.6145, 0.3607)&(-0.0397, 0.0159)&(-0.1067, 0.1067)\\
+1&-0.8&+1&-0.8&TR+LO&TR+LO&(-0.0870, 0.0801)&(-0.0099, 0.0262)&(-0.0590, 0.0590)\\
-1&+0.8&+1&-0.8&TR+LO&TR+LO&(-0.0863, 0.0794)&(-0.0261, 0.0089)&(-0.0630, 0.0630)\\
+1&-0.8&+1&-0.8&TR&TR+LO&(-0.0946, 0.0864)&(-0.0104, 0.0365)&(-0.0713, 0.0713)\\
+1&-0.8&+1&-0.8&LO&TR+LO&(-0.2482, 0.1978)&(-0.0168, 0.0161)&(-0.0606, 0.0606)\\
-1&+0.8&+1&-0.8&TR&TR+LO&(-0.1023, 0.0927)&(-0.0190, 0.0157)&(-0.0665, 0.0665)\\
-1&+0.8&+1&-0.8&LO&TR+LO&(-0.1726, 0.1469)&(-0.0512, 0.0070)&(-0.0946, 0.0946)\\
+1&-0.8&+1&-0.8&TR&LO&(-0.2869, 0.2213)&(-0.0230, 0.0199)&(-0.0875, 0.0875)\\
-1&+0.8&+1&-0.8&TR&LO&(-0.2247, 0.1828)&(-0.0487, 0.0089)&(-0.0924, 0.0924)\\
+1&-0.8&+1&-0.8&TR&TR&(-0.1020, 0.0925)&(-0.0103, 0.0492)&(-0.0790, 0.0790)\\
+1&-0.8&+1&-0.8&LO&LO&(-0.7443, 0.3909)&(-0.0152, 0.0172)&(-0.0545, 0.0545)\\
-1&+0.8&+1&-0.8&TR&TR&(-0.1181, 0.1056)&(-0.0136, 0.0289)&(-0.0717, 0.0717)\\
-1&+0.8&+1&-0.8&LO&LO&(-0.3029, 0.2305)&(-0.0624, 0.0098)&(-0.1749, 0.1749)\\
\end{tabular}
\end{ruledtabular}
\end{table}


\begin{thebibliography}{99}
\bibitem{barate} R. Barate \textit{et al.}, Phys. Lett. B 565, 61
(2003).
\bibitem{lep} LEP Electroweak Working Group, http://lepewwg.web.cern.ch/LEPEWWG/.
\bibitem{biswal} S. S. Biswal, D. Choudhury, R. M. Godbole, and R.
K. Singh, Phys. Rev. D 73, 035001 (2006).
\bibitem{garcia} M. C. Gonzalez-Garcia, Int.J.Mod.Phys. A14, 3121
(1999).
\bibitem{barger} V. Barger, T. Han, P. Langacker, B. McElrath and P.
Zerwas, Phys. Rev. D67, 115001 (2003).
\bibitem{hagiwara} S. Dutta, K. Hagiwara and Y. Matsumoto,
arXiv:0808.0477 [hep-ph].
\bibitem{lietti} M.C. Gonzalez-Garcia, S. M. Lietti and S. F. Novaes, Phys. Rev. D59, 075008 (1999).
\bibitem{choudhury} D. Choudhury and Mamta, Phys. Rev. D 74, 115019
(2006);\\\.{I}. \c{S}ahin,  Phys. Rev. D 77, 115010 (2008).
\bibitem{han} T. Han, Y-P. Kuang and B. Zhang, Phys. Rev. D 73, 055010 (2006).
\bibitem{he} H.-J. He, Y.-P. Kuang, C.-P. Yuan and B. Zhang, Phys. Lett. B 554, 64
(2003);\\
B. Zhang, Y.-P. Kuang, H.-J. He and C.-P. Yuan, Phys. Rev. D 67,
114024 (2003).
\bibitem{hankele} V. Hankele, G. Kl\"{a}mke, D. Zeppenfeld and T. Figy, Phys. Rev. D 74, 095001 (2006).
\bibitem{eboli} O. J. P. \'{E}boli, M. C. Gonzalez-Garcia and S.
F. Novaes, Phys. Rev. D 50, 3546 (1994);\\
O. J. P. \'{E}boli, M. C. Gonzalez-Garcia, F. Halzen and D.
Zeppenfeld, Phys. Rev. D 48, 1430 (1993);\\
J. F. Gunion and H. E. Haber Phys. Rev. D 48, 5109 (1993).
\bibitem{baillargeon} M. Baillargeon and F. Boudjema, Phys. Lett. B
317, 371 (1993).
\bibitem{boos} E. Boos, I. Ginzburg, K. Melnikov, T. Sack and S.
Shichanin, Z. Phys. C 56, 487 (1992);\\
K. Cheung, Phys. Rev. D 47, 3750 (1993).
\bibitem{Ginzburg} I.F. Ginzburg et al., Nucl. Instrum. Methods 205,
47 (1983);\\
I.F. Ginzburg et al., Nucl. Instrum. Methods 219, 5 (1984).
\bibitem{grace} T. Kaneko in ``New Computing Techniques in Physics
Research'', ed. D. Perret-Gallix, W. Wojcik (Paris: Edition du CNRS)
1990;\\
MINAMI-TATEYA group, ``GRACE manual'', KEK Report 92-19, 1993;\\ F.
Yuasa {\it et al.}, Prog. Theor. Phys. Suppl. 138, 18 (2000).
\bibitem{hagiwara2} K. Hagiwara et al., Nucl. Phys. B 282, 253 (1987).
\bibitem{abbiendi} G. Abbiendi et al. (OPAL Collaboration), Eur. Phys. J. C19, 229 (2001); Phys. Lett. B585,
223 (2004); P. Achard et al. (L3 Collaboration), Phys. Lett. B557,
147 (2003).
\end{thebibliography}
\end{document}